\documentclass[preprint,showpacs,preprintnumbers,amsmath,amssymb]{revtex4}
\usepackage{epsfig}
\usepackage{graphicx}
\usepackage{dcolumn}
\usepackage{bm}
\usepackage{threeparttable}

\def\beq{\begin{equation}}
\def\eeq{\end{equation}}
\def\eeqn{\end{equation}}
\newcommand\iden{\leavevmode\hbox{\small1\normalsize\kern-.33em1}}

\newcommand{\bea} {\begin{eqnarray}}
\newcommand{\eea} {\end{eqnarray}}

\let\jnfont=\rm
\def\NPB#1,{{\jnfont Nucl.\ Phys.\ B }{\bf #1},}
\def\PLB#1,{{\jnfont Phys.\ Lett.\ B }{\bf #1},}
\def\EPJC#1,{{\jnfont Eur.\ Phys.\ Jour.\ C }{\bf #1},}
\def\PRD#1,{{\jnfont Phys.\ Rev.\ D }{\bf #1},}
\def\PRL#1,{{\jnfont Phys.\ Rev.\ Lett.\ }{\bf #1},}
\def\MPLA#1,{{\jnfont Mod.\ Phys.\ Lett.\ A }{\bf #1},}
\def\JPG#1,{{\jnfont J.\ Phys.\ G }{\bf #1},}
\def\CTP#1,{{\jnfont Commun.\ Theor.\ Phys.\ }{\bf #1},}
\def\JHEP#1,{{\jnfont JHEP \ }{\bf #1},}
\def\NPPS#1,{{\jnfont Nucl.\ Phys.\ Proc.\ Suppl.\ }{\bf #1},}
\def\CPC#1,{{\jnfont Comput.\ Phys.\ Commun.\ }{\bf #1},}
\def\CPL#1,{{\jnfont Chin.\ Phys.\ Lett. }{\bf #1},}
\def\APPB#1,{{\jnfont Acta\ Phys.\ Polon.\ B }{\bf #1},}

\def\lsim{\raise0.3ex\hbox{$<$\kern-0.75em\raise-1.1ex\hbox{$\sim$}}}
\def\gsim{\raise0.3ex\hbox{$>$\kern-0.75em\raise-1.1ex\hbox{$\sim$}}}
\def\PR#1,{{\jnfont Phys.\ Rept. }{\bf #1},}
\def\CHC#1,{{\jnfont Chin.\ Phys.\ C }{\bf #1},}

\begin{document}

\title{\ \\[10mm] Two-Higgs-doublet model of type-II confronted with the LHC run-I and run-II data}

\author{Lei Wang, Xiao-Fang Han}
 \affiliation{Department of Physics, Yantai University, Yantai
264005, P. R. China}


\begin{abstract}
We examine the parameter space of two-Higgs-doublet model of type-II
after imposing the relevant theoretical and experimental constraints
from the precision electroweak data, $B$-meson decays, the LHC run-I
and run-II data. We find that the searches for Higgs via the
$\tau\bar{\tau},~WW,~ZZ,\gamma\gamma,~hh,~hZ,~HZ,~AZ$ channels can
give strong constraints on the CP-odd Higgs $A$ and heavy CP-even
Higgs $H$, and the parameter space excluded by each channel is
respectively carved out in detail. The surviving samples are
discussed in two different of regions: (i) In the SM-like coupling
region of the 125 GeV Higgs, $m_A$ is allowed to be as low as 350
GeV, and $\tan\beta$ is imposed a strong upper limit. $m_H$ is
allowed to be as low as 200 GeV for the proper $\tan\beta$,
$\sin(\beta-\alpha)$ and $m_A$, but is required to be larger than
300 GeV for $m_A=700$ GeV. (ii) In the wrong sign Yukawa coupling of
the 125 GeV, $\tan\beta$ and $\sin(\beta-\alpha)$ can be
 imposed upper limits by $b\bar{b}\to A/H\to \tau\bar{\tau}$ channel and lower limits by the
$A\to hZ$ channel. $m_A$ and $m_H$ are respectively allowed to be as low as 60 GeV and 200 GeV,
but 320 GeV $< m_A<$ 500 GeV is excluded for $m_H=$ 700 GeV.

\end{abstract}

\maketitle

\section{Introduction}
The ongoing analyses of ATLAS and CMS show that the properties of the newly discovered 125 GeV boson
are well consistent with the SM Higgs boson \cite{cmsh,atlh,160602266}. No excesses are observed in the searches
for the additional exotic Higgs. The ATLAS and CMS give us plentiful limits on additional scalar
state from its decay into various SM channels and some exotic decays at the LHC run-I and run-II.

The two-Higgs-doublet model (2HDM) \cite{2hdm} extends SM simply by
adding a second $SU(2)_L$ Higgs doublet, which has very rich Higgs
phenomenology, including two neutral CP-even Higgs bosons $h$ and
$H$, one neutral pseudoscalar $A$, and two charged Higgs $H^{\pm}$.
According to the different Yukawa couplings, there are four types
for 2HDMs which forbid the tree-level flavor changing neutral
currents, the type-I \cite{i-1,i-2}, type-II \cite{i-1,ii-2},
lepton-specific, and flipped models \cite{xy-1,xy-2,xy-3,xy-4}.
Since both the Yukawa couplings of down-type quark and lepton can be
enhanced by the factor of $\tan\beta$, the type-II model can be given
more stringent constraints than the other three models by the flavor
observables and the LHC searches for additional Higgs. The allowed
parameter space of 2HDM has been examined in light of the ATLAS and
CMS searches for extra Higgses at the LHC in the literatures
\cite{ph2h1,ph2h2,ph2h3,ph2h4,ph2h5,ph2h6,ph2h60,ph2h61,ph2h62,ph2h63,ph2h64,ph2h65,
ph2h7,ph2h8,ph2h9,ph2h11,ph2h12,mhp500,160104545,160401406}

In this paper we examine the
parameter space of the scenario in the type-II 2HDM considering the
joint constraints from the theory, the precision electroweak data,
flavor observables, the 125 GeV Higgs signal data and the searches
for the additional Higgs at the LHC run-I and run-II. The 2HDM can
respectively give the well fit to the 125 GeV Higgs signal data in
two different regions: wrong sign Yukawa coupling and SM-like
coupling. We respectively carve out the allowed
parameter space in the two different regions, and obtain some
interesting observables.

Our work is organized as follows. In Sec. II we introduce the
2HDM of type-II briefly. In Sec. III we perform
numerical calculations. In Sec. IV, we show the allowed parameter space
after imposing the relevant theoretical and
experimental constraints. Finally, we give our
conclusion in Sec. V.

\section{two-Higgs-doublet model of type-II}
The general Higgs potential with a softly broken discrete $Z_2$ symmetry is written as
\cite{2h-poten}
\begin{eqnarray} \label{V2HDM} \mathrm{V} &=& m_{11}^2
(\Phi_1^{\dagger} \Phi_1) + m_{22}^2 (\Phi_2^{\dagger}
\Phi_2) - \left[m_{12}^2 (\Phi_1^{\dagger} \Phi_2 + \rm h.c.)\right]\nonumber \\
&&+ \frac{\lambda_1}{2}  (\Phi_1^{\dagger} \Phi_1)^2 +
\frac{\lambda_2}{2} (\Phi_2^{\dagger} \Phi_2)^2 + \lambda_3
(\Phi_1^{\dagger} \Phi_1)(\Phi_2^{\dagger} \Phi_2) + \lambda_4
(\Phi_1^{\dagger}
\Phi_2)(\Phi_2^{\dagger} \Phi_1) \nonumber \\
&&+ \left[\frac{\lambda_5}{2} (\Phi_1^{\dagger} \Phi_2)^2 + \rm
h.c.\right].
\end{eqnarray}
We focus on the CP-conserving model in which all $\lambda_i$ and
$m_{12}^2$ are real.
The two complex scalar doublets have the hypercharge $Y = 1$:
\begin{equation}
\Phi_1=\left(\begin{array}{c} \phi_1^+ \\
\frac{1}{\sqrt{2}}\,(v_1+\phi_1^0+ia_1)
\end{array}\right)\,, \ \ \
\Phi_2=\left(\begin{array}{c} \phi_2^+ \\
\frac{1}{\sqrt{2}}\,(v_2+\phi_2^0+ia_2)
\end{array}\right).
\end{equation}
Where $v_1$ and $v_2$ are the electroweak vacuum expectation values
(VEVs) with $v^2 = v^2_1 + v^2_2 = (246~\rm GeV)^2$, and the ratio of the two VEVs is defined
as $\tan\beta=v_2 /v_1$. After spontaneous electroweak
symmetry breaking, there are five mass eigenstates: two neutral
CP-even $h$ and $H$, one neutral pseudoscalar $A$, and two charged
scalars $H^{\pm}$.

The Yukawa interactions are written as
 \bea
- {\cal L} &=&Y_{u2}\,\overline{Q}_L \, \tilde{{ \Phi}}_2 \,u_R
+\,Y_{d1}\,
\overline{Q}_L\,{\Phi}_1 \, d_R\, + \, Y_{\ell 1}\,\overline{L}_L \, {\Phi}_1\,e_R+\, \mbox{h.c.}\,, \eea where
$Q_L^T=(u_L\,,d_L)$, $L_L^T=(\nu_L\,,l_L)$,
$\widetilde\Phi_{1,2}=i\tau_2 \Phi_{1,2}^*$, and $Y_{u2}$,
$Y_{d1}$ and $Y_{\ell 1}$ are $3 \times 3$ matrices in family
space.

The Yukawa couplings of the neutral Higgs bosons normalized to the SM are given as
\bea\label{hffcoupling} &&
y_{h}^{f_i}=\left[\sin(\beta-\alpha)+\cos(\beta-\alpha)\kappa_f\right], \nonumber\\
&&y_{H}^{f_i}=\left[\cos(\beta-\alpha)-\sin(\beta-\alpha)\kappa_f\right], \nonumber\\
&&y_{A}^{f_i}=-i\kappa_f~{\rm (for~u)},~~~~y_{A}^{f_i}=i \kappa_f~{\rm (for~d,~\ell)}, \eea where
$\kappa_d=\kappa_\ell\equiv-\tan\beta$ and $\kappa_u=1/\tan\beta$.

The Yukawa interactions of the charged Higgs are given as
\begin{align} \label{eq:Yukawa2}
 \mathcal{L}_Y & = - \frac{\sqrt{2}}{v}\, H^+\, \Big\{\bar{u}_i \left[\kappa_d\,(V_{CKM})_{ij}~ m_{dj} P_R
 - \kappa_u\,m_{ui}~ (V_{CKM})_{ij} ~P_L\right] d_j + \kappa_\ell\,\bar{\nu} m_\ell P_R \ell
 \Big\}+h.c.,
 \end{align}
where $i,j=1,2,3$.

The neutral Higgs boson couplings with the gauge bosons normalized to the
SM are given by
\beq
y^{V}_h=\sin(\beta-\alpha),~~~
y^{V}_H=\cos(\beta-\alpha),\label{hvvcoupling}
\eeq
where $V$ denotes $Z$ or $W$.

The properties of the observed 125 GeV Higgs are very closed to the
SM Higgs boson, which gives the strong constraints on the sector of
Higgs extensions. The 2HDM can give the well fit to the 125 GeV
Higgs signal data in two different regions: the SM-like coupling and
the wrong sign Yukawa coupling of 125 GeV Higgs. For the former, the
couplings of 125 GeV Higgs are very closed to the SM Higgs, which
has a limit called the alignment limit. In the exact alignment limit
\cite{ph2h9,alignment2}, namely $\cos(\beta-\alpha)=0$, from Eq.
(\ref{hffcoupling}) and Eq. (\ref{hvvcoupling}) we find that $h$ has
the same couplings to the fermions and gauge bosons as the SM, and
the heavy CP-even Higgs ($H$) has no couplings to the gauge bosons.

In the wrong sign Yukawa coupling region, at least one of the Yukawa couplings of the 125 GeV Higgs
has the opposite sign to the coupling of gauge boson. However, their absolute values should be closed to the SM Higgs
due to the constraints of 125 GeV Higgs signal data. Therefore, we can obtain
\begin{align}  &y_h^{f_i}=-1+\epsilon,~~y^{V}_h\simeq 1-0.5\cos^2(\beta-\alpha) ~~{\rm for}~ \sin(\beta-\alpha) >0~{\rm and}~\cos(\beta-\alpha) >0~,\nonumber\\
& y_h^{f_i}=1-\epsilon,~~y^{V}_h\simeq -1+0.5\cos^2(\beta-\alpha) ~~{\rm for}~ \sin(\beta-\alpha)<0~{\rm and}~\cos(\beta-\alpha) >0. \end{align}
Where $\mid\epsilon\mid$ and $\mid\cos(\beta-\alpha)\mid$ are much smaller than 1.
From Eq. (\ref{hffcoupling}), we can obtain
\begin{align}\label{wrcp}
&\kappa_f=\frac{-2+\varepsilon+0.5\cos(\beta-\alpha)^2}{\cos(\beta-\alpha)}<<-1 ~{\rm for}~ \sin(\beta-\alpha) >0~{\rm and}~\cos(\beta-\alpha) >0~,\nonumber\\
&\kappa_f=\frac{2-\varepsilon-0.5\cos(\beta-\alpha)^2}{\cos(\beta-\alpha)} >>1 ~{\rm for}~ \sin(\beta-\alpha) <0~{\rm and}~\cos(\beta-\alpha) >0~.
\end{align}
Therefore, in the 2HDM of type-II there are the wrong sign Yukawa couplings of the down-type quark and lepton only
for $\sin(\beta-\alpha) > 0$ and $\cos(\beta-\alpha) >0$ since $\tan\beta$ is required to be larger than 1 by the
$B$-meson observables and $R_b$. For the same $\sin(\beta-\alpha)$, especially for $\sin(\beta-\alpha)\to 1$, $\tan\beta$
in the wrong sign Yukawa coupling region is much larger than that of the SM-like coupling region. In other words,
$\tan\beta$ has a lower bound in the wrong sign Yukawa coupling region and is allowed to be as low as 1 in the SM-like coupling.
In addition, $\cos(\beta-\alpha)$ in the wrong sign Yukawa coupling is allowed to be much larger than that of the SM-like
coupling due to the presence of "-2" of the numerator in the Eq. (\ref{wrcp}).

\section{Numerical calculations}
We take the light CP-even Higgs boson $h$ as the
SM-like Higgs, $m_h=125$ GeV. The measurement of the branching fraction of $b \to s\gamma$ gives the
stringent constraints on the charged Higgs mass of 2HDM of type-II, $m_{H^{\pm}} > 570$ GeV \cite{bsr570}.

In our calculation, we consider the following observables and constraints:

\begin{itemize}
\item[(1)] Theoretical constraints and precision electroweak data. The $\textsf{2HDMC}$ \cite{2hc-1,2hc-2}
is employed to implement the theoretical
constraints from the vacuum stability, unitarity and
coupling-constant perturbativity, as well as the constraints from
the oblique parameters ($S$, $T$, $U$).

\item[(2)] The flavor observables and $R_b$. We consider the constraints of $B$-meson
decays from $B\to X_s\gamma$, $\Delta m_{B_s}$ and $\Delta m_{B_d}$.  $\textsf{SuperIso-3.4}$ \cite{spriso} is
used to calculate $B\to X_s\gamma$, and $\Delta m_{B_s}$ and $\Delta m_{B_d}$ are respectively calculated using the
formulas in \cite{deltmq}. In addition, we consider the constraints of bottom quarks produced in $Z$ decays, $R_b$,
which is calculated following the formulas in \cite{rb1,rb2}.

\begin{table}
\begin{footnotesize}
\begin{tabular}{| c | c | c | c |}
\hline
\textbf{Channel} & \textbf{Experiment} & \textbf{Mass range (GeV)}  &  \textbf{Luminosity} \\
\hline
 {$gg/b\bar{b}\to H/A \to \tau^{+}\tau^{-}$} & ATLAS 8 TeV~\cite{47Aad:2014vgg} & 90-1000 & 19.5-20.3 fb$^{-1}$ \\
{$gg/b\bar{b}\to H/A \to \tau^{+}\tau^{-}$} & CMS 8 TeV~\cite{48CMS:2015mca} &  90-1000  &19.7 fb$^{-1}$ \\
 {$b\bar{b}\to H/A \to \tau^{+}\tau^{-}$} & CMS 8 TeV \cite{1511.03610}& 20-80   & 19.7 fb$^{-1}$ \\
{$gg/b\bar{b}\to H/A \to \tau^{+}\tau^{-}$} & ATLAS 13 TeV~\cite{82vickey} & 200-1200 &13.3 fb$^{-1}$ \\
\hline

 {$pp\to H/A \to \gamma\gamma$} & ATLAS 13 TeV \cite{80lenzi} & 200-2400 & 15.4 fb$^{-1}$ \\
{$gg\to H/A \to \gamma\gamma$}& CMS 8+13 TeV \cite{81rovelli}& 500-4000 & 12.9 fb$^{-1}$ \\
\hline

 {$gg/VV\to H\to W^{+}W^{-}$} & ATLAS 8 TeV  \cite{55Aad:2015agg}& 300-1500  &  20.3 fb$^{-1}$\\

{$gg/VV\to H\to W^{+}W^{-}~(\ell\nu\ell\nu)$} & ATLAS 13 TeV  \cite{77atlasww13}& 300-3000  &  13.2 fb$^{-1}$\\

{$gg\to H\to W^{+}W^{-}~(\ell\nu qq)$} & ATLAS 13 TeV  \cite{78atlasww13lvqq}& 500-3000  &  13.2 fb$^{-1}$\\
\hline

$gg/VV\to H\to ZZ$ & ATLAS 8 TeV \cite{57Aad:2015kna}& 160-1000 & 20.3 fb$^{-1}$ \\

$gg\to H \to ZZ(\ell \ell \nu \nu)$ & ATLAS 13 TeV~\cite{74koeneke4} & 300-1000  & 13.3 fb$^{-1}$ \\
$gg\to H\to ZZ(\nu \nu qq)$ & ATLAS 13 TeV~\cite{75koeneke5} & 300-3000 & 13.2 fb$^{-1}$ \\
$gg/VV\to H\to ZZ(\ell \ell qq)$ & ATLAS 13 TeV~\cite{75koeneke5} & 300-3000 & 13.2 fb$^{-1}$ \\
$gg/VV\to H\to ZZ(4\ell)$ & ATLAS 13 TeV~\cite{76koeneke3} & 200-3000 & 14.8 fb$^{-1}$ \\
\hline

$gg\to H\to hh \to (\gamma \gamma) (b \bar{b})$ & CMS 8 TeV \cite{64Khachatryan:2016sey} & 250-1100  & 19.7 fb$^{-1}$\\

$gg\to H\to hh \to (b\bar{b}) (b\bar{b})$ & CMS 8 TeV \cite{65Khachatryan:2015yea}&   270-1100   & 17.9 fb$^{-1}$\\

$gg\to H\to hh \to (b\bar{b}) (\tau^{+}\tau^{-})$ & CMS 8 TeV \cite{66Khachatryan:2015tha}&  260-350 & 19.7 fb$^{-1}$\\

$gg \to H\to hh \to b\bar{b}b\bar{b}$ & ATLAS 13 TeV~\cite{84varol} & 300-3000  &  13.3 fb$^{-1}$ \\

$gg \to H\to hh \to b\bar{b} \tau^{+} \tau^{-}$ & CMS 13 TeV~\cite{85CMS:2016knm} & 250-900  &  12.9 fb$^{-1}$ \\
\hline

$gg\to A\to hZ \to (\tau^{+}\tau^{-}) (\ell \ell)$ & CMS 8 TeV \cite{66Khachatryan:2015tha}& 220-350 & 19.7 fb$^{-1}$\\

$gg\to A\to hZ \to (b\bar{b}) (\ell \ell)$ & CMS 8 TeV \cite{67Khachatryan:2015lba} & 225-600 &19.7 fb$^{-1}$ \\

$gg\to A\to hZ\to (\tau^{+}\tau^{-}) Z$ & ATLAS 8 TeV \cite{68Aad:2015wra}&220-1000 & 20.3 fb$^{-1}$ \\

 {$gg\to A\to hZ\to (b\bar{b})Z$} & ATLAS 8 TeV \cite{68Aad:2015wra}& 220-1000 & 20.3 fb$^{-1}$  \\
{$gg/b\bar{b}\to A\to hZ\to (b\bar{b})Z$}& ATLAS 13 TeV \cite{69AZhatlas13}& 200-2000 & 3.2 fb$^{-1}$  \\
\hline

$gg\to A(H)\to H(A)Z\to (b\bar{b}) (\ell \ell)$ & CMS 8 TeV \cite{160302991} & 200-1000 &19.8 fb$^{-1}$ \\

$gg\to A(H)\to H(A)Z\to (\tau^{+}\tau^{-}) (\ell \ell)$ & CMS 8 TeV \cite{160302991}& 200-1000 & 19.8 fb$^{-1}$ \\

$gg\to A(H)\to H(A)Z\to (b\bar{b}) (\ell \ell)$ & CMS 13 TeV \cite{HIG-16-010}& 50-800 & 2.3 fb$^{-1}$ \\
\hline

\end{tabular}
\end{footnotesize}
\caption{The upper limits at 95\%  C.L. on the production cross-section times branching ratio of the processes considered in the $ H $ and $ A $  searches at the LHC run-I and run-II.}
\label{tabh}
\end{table}

\item[(3)] The global fit to the 125 GeV Higgs signal data. We perform the
$\chi^2$ calculation for the signal strengths of the 125 GeV Higgs in the
$\mu_{ggF+tth}(Y)$ and $\mu_{VBF+Vh}(Y)$ with $Y$ denoting the decay
mode $\gamma\gamma$, $ZZ$, $WW$, $\tau\bar{\tau}$ and $b\bar{b}$,
 \begin{eqnarray} \label{eq:ellipse}
  \chi^2(Y) =\left( \begin{array}{c}
        \mu_{ggH+ttH}(Y) - \widehat{\mu}_{ggH+ttH}(Y)\\
        \mu_{VBF+VH}(Y) - \widehat{\mu}_{VBF+VH}(Y)
                 \end{array} \right)^T
                 \left(\begin{array}{c c}
                        a_Y & b_Y \\
                        b_Y & c_Y
                 \end{array}\right) \nonumber\\
\times
                  \left( \begin{array}{c}
        \mu_{ggH+ttH}(Y) - \widehat{\mu}_{ggH+ttH}(Y)\\
        \mu_{VBF+VH}(Y) - \widehat{\mu}_{VBF+VH}(Y)
                 \end{array} \right) \,.
 \end{eqnarray}
$\widehat{\mu}_{ggH+ttH}(Y)$ and $\widehat{\mu}_{VBF+VH}(Y)$
are the data best-fit values and $a_Y$, $b_Y$ and $c_Y$ are the
parameters of the ellipse, which are given by the
combined ATLAS and CMS experiments \cite{160602266}. We pay
particular attention to the surviving samples with
$\chi^2-\chi^2_{\rm min} \leq 6.18$, where $\chi^2_{\rm min}$
denotes the minimum of $\chi^2$. These samples correspond to be within
the $2\sigma$ range in any two-dimension plane of the
model parameters when explaining the Higgs data.

\item[(4)] The non-observation of additional Higgs bosons. We employ
$\textsf{HiggsBounds-4.3.1}$ \cite{hb1,hb2} to implement the exclusion
constraints from the neutral and charged Higgs searches at LEP at 95\% confidence level.

At the LHC run-I and run-II the ATLAS and CMS have searched for the
additional scalar state from its decay into various SM channels and
some exotic decays. For the $gg\to A$ production in 2HDM of type-II,
there is the destructive interference contributions of $b$-quark
loop and top quark loop. The cross section decreases with increasing
of $\tan\beta$, reaches the minimum value for the moderate value of
$\tan\beta$, and is dominated by the $b$-quark loop for enough large
value of $\tan\beta$. For the $gg\to H$ production, the cross
section depends on $\sin(\beta-\alpha)$ in addition to $\tan\beta$
and $m_H$. We compute the cross sections for $H$ and $A$ in the
gluon fusion and $b\bar{b}$-associated production at NNLO in QCD via
$\textsf{SusHi}$ \cite{sushi}. The production cross sections of $H$
in vector boson fusion process are taken from results of the LHC
Higgs Cross Section Working Group \cite{higgswg}. The
$\textsf{2HDMC}$ is used to calculate the branching ratios of the
various decay modes of $H$ and $A$. A complete list of the
additional Higgs searches considered by us is summarized in the
Table \ref{tabh}. For 1$\leq \tan\beta \leq 30$, the LHC searches
for the heavy charged scalar can not give the constraints on the
model for $m_{H^{\pm}}>500$ GeV \cite{mhp500}. Therefore, we do not
include the searches for the heavy charged Higgs.
\end{itemize}

\section{Results and discussions}
\subsection{The constraints from the 125 GeV Higgs signal data and the oblique parameters}
\begin{figure}[tb]
 \epsfig{file=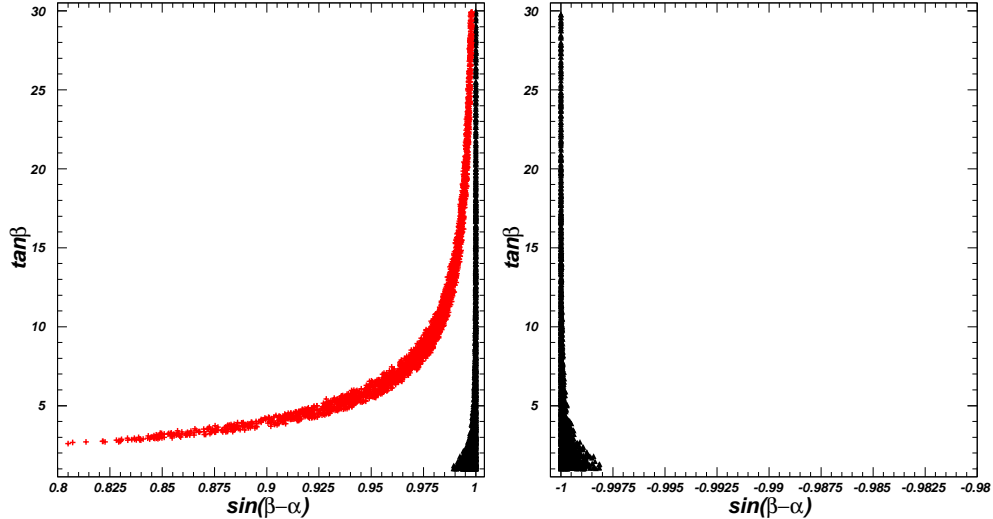,height=7.0cm}
\vspace{-0.2cm} \caption{The samples surviving from the constraints of the 125 GeV Higgs signal data
projected on the plane of $\sin(\beta-\alpha)$ versus $\tan\beta$.} \label{athe1}
\end{figure}

The couplings of SM-like Higgs are sensitive to the parameters $\sin(\beta-\alpha)$ and $\tan\beta$.
Therefore, the signal data of the 125 GeV Higgs can give the strong constraints on the two parameters.
In Fig. \ref{athe1}, we show $\sin(\beta-\alpha)$ and $\tan\beta$ allowed by the the signal data of the
125 GeV Higgs. From Fig. \ref{athe1}, we find that in the SM-like coupling region of the 125 GeV Higgs,
$\mid\sin(\beta-\alpha)\mid$ is required to be within the very narrow ranges, $\mid\sin(\beta-\alpha)\mid>0.99$.
However, in the wrong sign Yukawa coupling region of the 125 GeV Higgs, $\sin(\beta-\alpha)$ is allowed to be much smaller than
1, but is always positive. $\tan\beta$ is imposed a lower bound for
a given value of $\sin(\beta-\alpha)$ in the wrong sign Yukwawa coupling region, such as $\tan\beta>3$ (7) for
$\sin(\beta-\alpha)=0.87~(0.97)$, while $\tan\beta$ is allowed to be as low as 1 for any value of $\sin(\beta-\alpha)$
in the SM-like Higgs coupling region.

\begin{figure}[tb]
 \epsfig{file=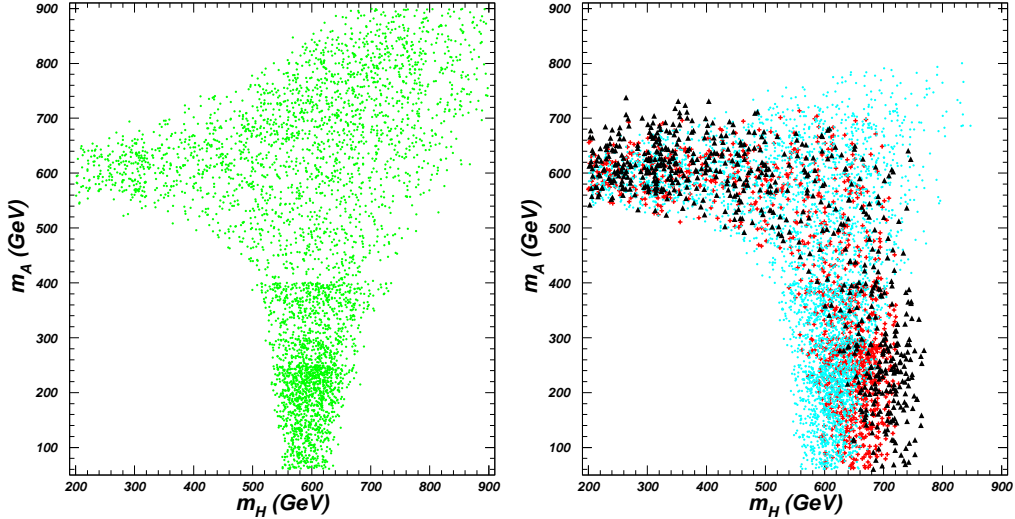,height=7.0cm}
\vspace{-0.3cm} \caption{Under the constraints of vacuum stability, unitarity, perturbativity, and oblique parameters,
the surviving samples projected on the planes of $m_H$ versus $m_A$.
Left panel: $0.99\leq\mid\sin(\beta-\alpha)\mid\leq 1$ for all the samples; Right panel: $0.95\leq\sin(\beta-\alpha)\leq 0.99$
for the bullets (sky blue), $0.90\leq\sin(\beta-\alpha)\leq 0.95$ for the pluses (red), and
 $0.80\leq\sin(\beta-\alpha)\leq 0.90$ for the triangles (black).} \label{athe2}
\end{figure}

The oblique parameters $S$, $T$ and $U$ can impose the strong constraints on the
2HDM mass spectrum. In Fig. \ref{athe2},
we show $m_H$ and $m_A$ allowed by the constraints of theory and oblique parameters.
The other relevant free parameters are scanned  in the following range:
 \begin{align}
&~~~~~~~~~~ 0.8\leq \mid\sin(\beta-\alpha)\mid \leq 1,~~ 1\leq \tan\beta \leq 30,
\nonumber\\
&570 {\rm\  GeV} \leq ~m_{H^{\pm}}  \leq 900  {\rm\  GeV},~~-(3000~{\rm GeV})^2 \leq m_{12}^2 \leq\  (3000~{\rm GeV})^2.
\label{scan}\end{align}

Fig. \ref{athe2} shows that at least one of $A$ and $H$ is required to have a large mass. It is disfavored that
both $m_A$ and $m_H$ are smaller than 440 GeV. A light $A$ ($H$) favors $m_H$ ($m_A$)
to be around 600 GeV for 0.99 $<\mid\sin(\beta-\alpha)\mid<$ 1.
In such range of $\sin(\beta-\alpha)$, the 125 GeV Higgs is allowed to have the SM-like coupling.
With the decreasing of $\mid\sin(\beta-\alpha)\mid$, a light $A$ favors $m_H$ to increase.
For example, for 0.8 $<\sin(\beta-\alpha) < 0.9$, a light $A$ favors $m_H$ to be around 700 GeV.
In such range of $\sin(\beta-\alpha)$, the 125 GeV Higgs is required to have the wrong-sign Yukawa coupling.

Now we carve out the allowed parameter space after imposing the joint constraints from the theory, the precision electroweak data,
flavor observables, the 125 GeV Higgs signal data, and especially for the searches for the additional Higgs at the LHC run-I and run-II.
The free parameters $\sin(\beta-\alpha)$, $\tan\beta$, $m_{12}^2$, and $m_{H^{\pm}}$ are scanned in the ranges shown
in Eq. (\ref{scan}). If at the same time we scan $m_A$ and $m_H$ randomly,
the constraints of each Higgs search channel on the property of $H$ or $A$ are not explicitly examined.
Therefore, in each analysis we will fix one of $m_A$ and $m_H$, and perform a detailed examination on the constraints of
the Higgs searches channels on another Higgs.
In light of the allowed Higgs mass spectrum shown in Fig. \ref{athe2}, we take four cases:
(a) $m_H=$ 600 GeV; (b) $m_H=$ 700 GeV; (a) $m_A=$ 600 GeV; (b) $m_A=$ 700 GeV.
For the four cases another Higgs is allowed to have a wide mass range including
the low mass. Since a heavy Higgs can easily avoid the constraints of the Higgs searches, the light Higgs is more interesting.

\subsection{Constraints on the CP-odd Higgs}
\begin{figure}[tb]
 \epsfig{file=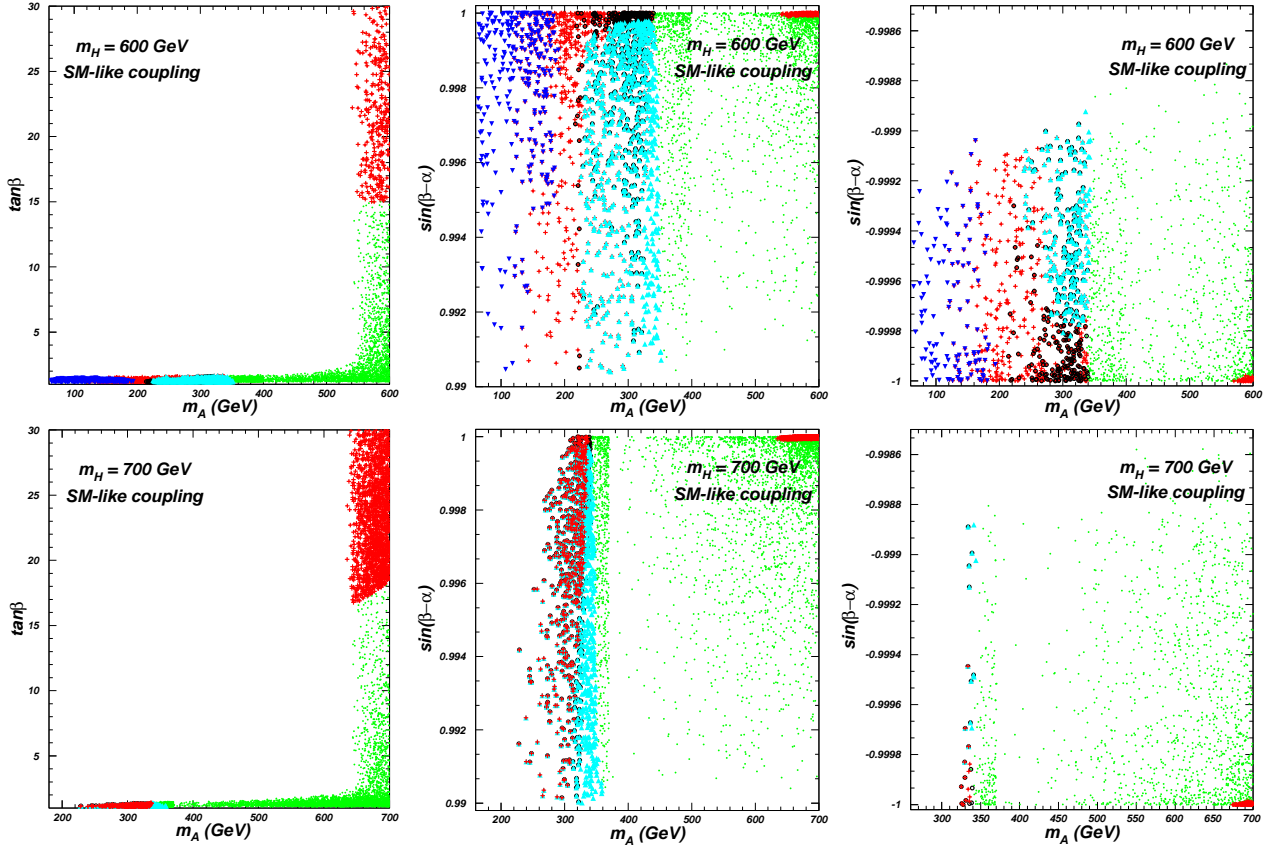,height=11.3cm}
\vspace{-1.0cm} \caption{The surviving samples of the
SM-like coupling region projected on the planes of $m_A$ versus
$\tan\beta$ and $m_A$ versus $\sin(\beta-\alpha)$. All the samples
are allowed by the constraints of pre-LHC (denoting theoretical
constrains, electroweak precision data, the flavor observables,
$R_b$, the exclusion limits from searches for Higgs at LEP) and the
125 GeV Higgs signal data. The pluses (red), circles (black),
triangles (sky blue) and inverted triangles (royal blue) are
respectively excluded by the $A/H\to \tau\bar{\tau}$, $A\to
\gamma\gamma$, $A\to hZ$ and $A\to HZ$ searches at the LHC
run-I and run-II.} \label{asm}
\end{figure}
In Fig. \ref{asm}, fixing $m_H=600$ GeV and $m_H=700$ GeV, we project the surviving samples of the SM-like
coupling region on the planes of $m_A$ versus $\tan\beta$ and $m_A$
versus $\sin(\beta-\alpha)$ after imposing the constraints of
pre-LHC (denoting theoretical constrains, electroweak precision
data, the flavor observables, $R_b$, the exclusion limits from
searches for Higgs at LEP), the 125 GeV Higgs signal data, and the
searches for additional Higgses at the LHC run-I and run-II.
The upper-left panel shows that the constraints of pre-LHC and the 125 GeV Higgs data
give the strong constraints on $\tan\beta$ and $m_A$.
$\tan\beta$ is required to be smaller than 2 for $m_A<$ 500 GeV and
$m_H=600$ GeV. Further the $A\to
\tau\bar{\tau}$, $A\to \gamma\gamma$, $A\to hZ$ and $H\to AZ$ channels
exclude the samples in the region of $m_A<350$ GeV.
The $b\bar{b}\to A/H\to \tau\bar{\tau}$ channel can give the upper
limit of $\tan\beta$, $\tan\beta<15$ for $m_A$ $(m_H)$ $<$ 600 GeV. In the SM-like Higgs coupling region,
the coupling constant of $Hb\bar{b}$ is nearly the same as that of $Ab\bar{b}$.
In our analysis we fix $m_H$ = 600 GeV which can lead to $\tan\beta<15$ due to the constraints of $b\bar{b}\to H\to \tau\bar{\tau}$ channel.
Therefore, we do not show the parameter space $m_A>600$ GeV although the constraints of
$b\bar{b}\to A\to \tau\bar{\tau}$ on $\tan\beta$ can be relaxed.

The upper-middle and upper-right panels show that the $H\to AZ$,
 $A\to \gamma\gamma$, $A\to hZ$ and $A\to\tau\bar{\tau}$ searches can give the strong constraints on
$m_A$ in the closed to alignment limit. With the increasing of $\mid\sin(\beta-\alpha)\mid$,
the widths of $H\to AZ$ and $A\to hZ$ increase and decrease, respectively.
The constraints from the $H\to AZ$ channel become strong as $\mid\sin(\beta-\alpha)\mid$ approaches to 1, and can
exclude most of samples in the range of $m_A<$ 200 GeV.
The $A\to hZ$ channel can exclude most of samples in the range of 220 GeV $<m_A<$ 350 GeV
except for the samples in the very closed to the alignment limit. For 350 GeV $<m_A<$ 540 GeV, the $A\to t\bar{t}$ can enhance
the total width of $A$ sizably, therefore the constraints of $H\to AZ$, $A\to \gamma\gamma$ and $A\to hZ$ channels can be satisfied.
For 540 GeV $<m_A<$ 600 GeV, some samples in the very closed to the alignment limit can be excluded by the $b\bar{b}\to A/H\to \tau\bar{\tau}$ channel (also see the upper-left panel).

For $m_H=700$ GeV, the constraints of pre-LHC and the 125 GeV Higgs data require $m_A$
to be larger than 220 GeV, and $\tan\beta$ to be smaller than 2 for $m_A<$ 640 GeV.
The $H\to AZ$ channel can not impose the further constraints on the parameter space.
The $A\to \tau\bar{\tau}$, $A\to hZ$ and $A\to \gamma\gamma$ channels can exclude $m_A < 340$ GeV.
350 GeV $<m_A<$ 640 GeV are allowed by the constraints of pre-LHC, the 125 GeV Higgs data and the Higgs searches at the LHC.
For 640 GeV $<m_A<$ 700 GeV, the $b\bar{b}\to A \to \tau\bar{\tau}$ can exclude the exact alignment limit and
require $\tan\beta$ to be smaller than 18.

\begin{figure}[tb]
 \epsfig{file=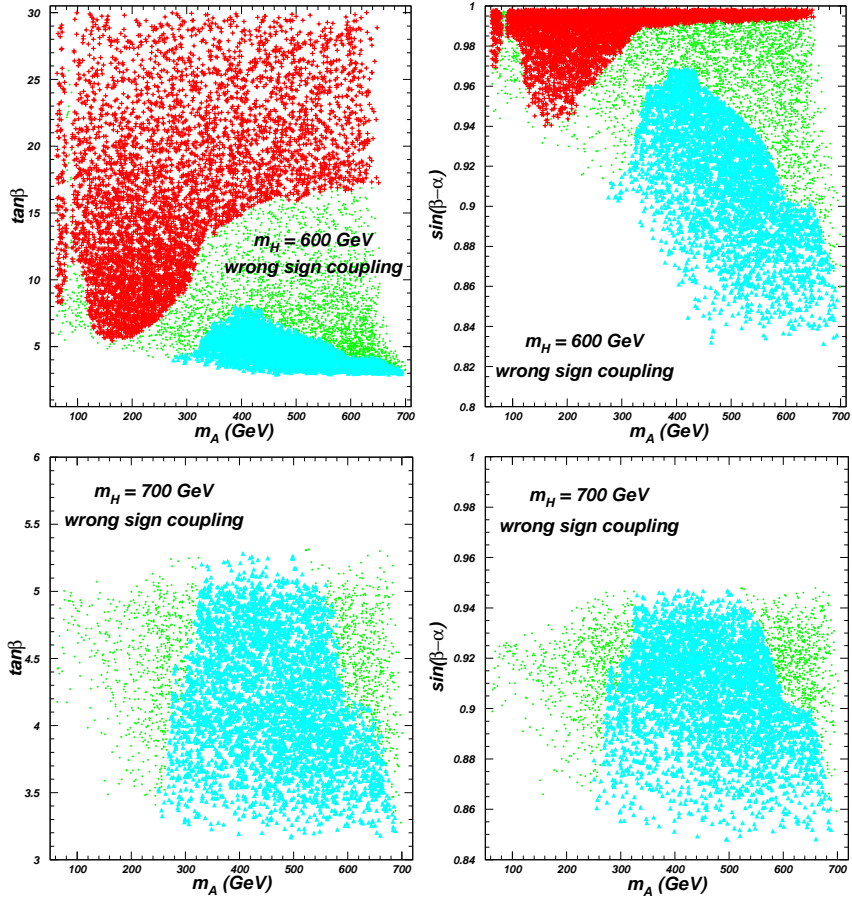,height=12cm}
\vspace{-0.5cm} \caption{The surviving samples of the
wrong sign Yukawa coupling region projected on the planes of $m_A$ versus
$\tan\beta$ and $m_A$ versus $\sin(\beta-\alpha)$. All the samples
are allowed by the constraints of pre-LHC and the
125 GeV Higgs signal data. The pluses (red) and
triangles (sky blue) are
respectively excluded by the $A/H\to \tau\bar{\tau}$ and $A\to hZ$ searches at the LHC
run-I and run-II.} \label{aw}
\end{figure}

Now we examine the parameter space in the wrong sign Yukawa coupling region for $m_H=$ 600 GeV and $m_H=$ 700 GeV.
In Fig. \ref{aw}, we project the
surviving samples on the planes of $m_A$ versus $\tan\beta$ and $m_A$ versus $\sin(\beta-\alpha)$
after imposing the constraints of pre-LHC, the 125 GeV Higgs signal data and
the searches for additional Higgses at the LHC run-I and run-II. Since the constraints of pre-LHC and the 125 GeV Higgs
signal data require $\tan\beta>3$ in the wrong sign Yukawa coupling region, the $A\to \gamma\gamma$
and $A\to HZ$ can not give the further constraints on the parameter space.

For $m_H=$ 600 GeV and 280 GeV $<m_A<700$ GeV, $\tan\beta$ and $\sin(\beta-\alpha)$ can be imposed the
upper bounds by the $b\bar{b}\to A \to \tau\bar{\tau}$ channel
and the lower bounds by the $A\to hZ$ channel. For example, 4.5 $<\tan\beta<$ 9.0 and 0.91$<\sin(\beta-\alpha)<$ 0.975
for $m_A=$ 300 GeV, 8.0 $<\tan\beta<$ 15.0 and 0.97$<\sin(\beta-\alpha)<$ 0.99
for $m_A=$ 400 GeV, 4.0 $<\tan\beta<$ 17.0 and 0.9$<\sin(\beta-\alpha)<$ 0.99
for $m_A=$ 600 GeV. The $b\bar{b}\to A \to \tau\bar{\tau}$ channel can exclude most of samples in the range of
 $m_A<$ 200 GeV except for a very narrow band of $m_A$ around 100 GeV.

For $m_H=700$ GeV, the constraints of pre-LHC and the 125 GeV Higgs
signal data require $\tan\beta<5.5$ and $\sin(\beta-\alpha)<$ 0.95. Therefore,
the $b\bar{b}\to A \to \tau\bar{\tau}$ channel can not give the further constraints on the parameter space.
The $A \to hZ$ channel can exclude most of samples in the range of 300 GeV $<m_A<600$ GeV.

\begin{figure}[tb]
 \epsfig{file=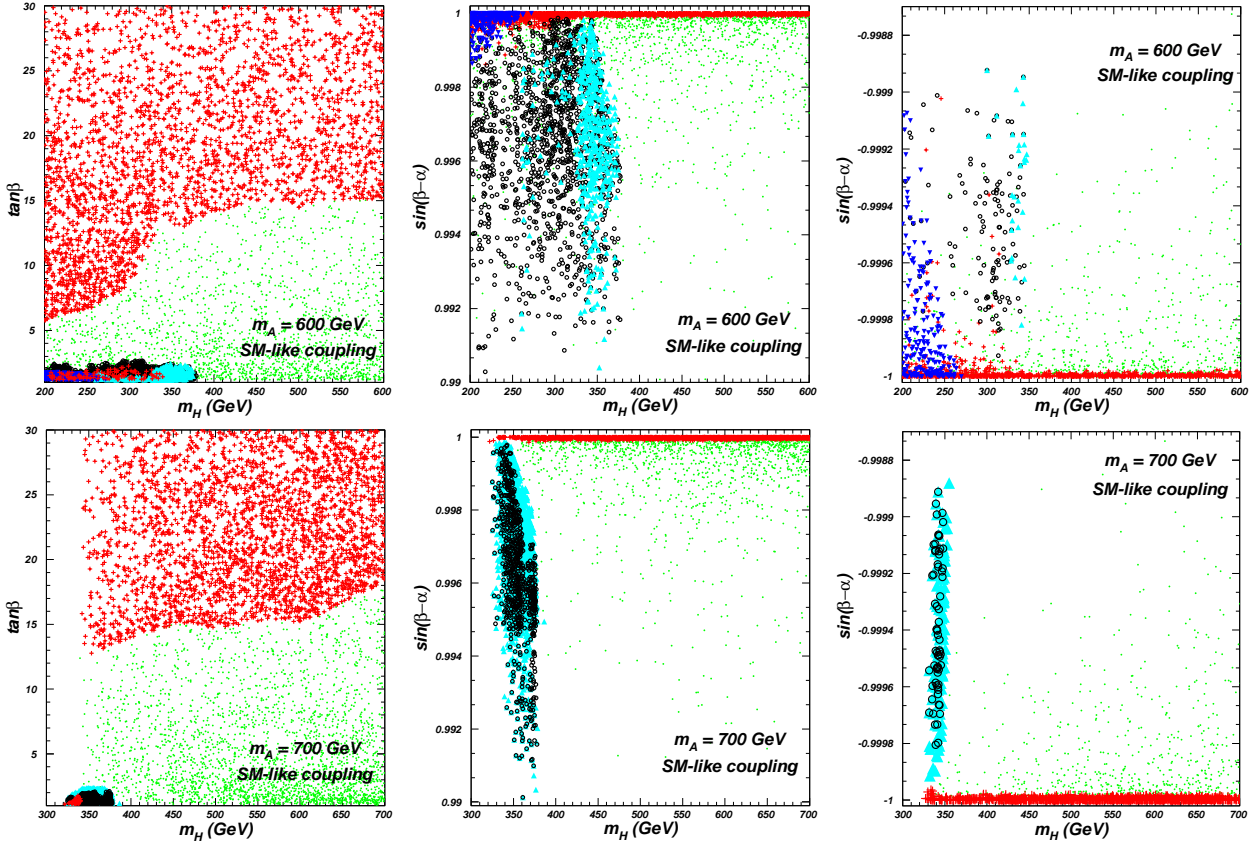,height=11.3cm}
\vspace{-0.5cm} \caption{The surviving samples of the
SM-like coupling region projected on the planes of $m_A$ versus
$\tan\beta$ and $m_A$ versus $\sin(\beta-\alpha)$. All the samples
are allowed by the constraints of pre-LHC and the
125 GeV Higgs signal data. The pluses (red), circles (black),
triangles (sky blue) and inverted triangles (royal blue) are
respectively excluded by the $H/A\to \tau\bar{\tau}$, $H\to
WW,~ZZ,~\gamma\gamma$, $H\to hh$, and $A\to HZ$ searches at the LHC
run-I and run-II.} \label{bsm}
\end{figure}

\subsection{Constraints on the heavy CP-even Higgs}
Here we examine the status of the heavy CP-even Higgs after imposing the relevant theoretical and experimental constraints.
In Fig. \ref{bsm}, fixing $m_A=600$ GeV and $m_A=$ 700 GeV, we project the
surviving samples on the planes of $m_H$ versus $\tan\beta$ and $m_H$ versus $\sin(\beta-\alpha)$.
For $m_A=$ 600 GeV, the upper-left panel shows the $b\bar{b}\to H/A\to
\tau\bar{\tau}$ channel gives the upper bound of $\tan\beta$, such
as $\tan\beta<6$, 10 and 15 for $m_H=200$ GeV, 320 GeV and 600 GeV.
The $H\to \tau\bar{\tau},~VV,~\gamma\gamma,~hh$ and $A\to HZ$
searches can require $\tan\beta>$ 2.5 for $m_H<$ 380 GeV.
The proper large $\tan\beta$ can simultaneously suppress the cross sections of Higgs in the gluon fusion production
and $b\bar{b}$-quark associated production. In addition, the $H\to b\bar{b}$ can be the dominant decay mode, so that
the branching ratios of $H\to VV,~\gamma\gamma,~hh$ will be sizably suppressed, and the constraints from
these channels can be avoided.

For $m_A=$ 600 GeV, the upper-middle and upper-right panels of Fig.
\ref{bsm} show for the proper $\tan\beta$ (see the upper-left
panel), the $H/A\to \tau\bar{\tau}$, $H\to VV,~\gamma\gamma,~hh$,
$A\to HZ$ can give the strong constraints on the $m_H$ in the closed
to the alignment limit. All the samples in the range of $m_H>$ 270
GeV can satisfy the constraints of $A\to HZ$ channel. The widths of
$H\to VV,~hh$ decrease with increasing of
$\mid\sin(\beta-\alpha)\mid$ and equal to zero in the exact
alignment limit. In the proper deviation from the alignment limit,
the searches for these channels can exclude most of samples in the
range of $m_H<$ 380 GeV. With the increasing of $m_H$, the $H\to
t\bar{t}$ can enhance the total width of $H$ sizably, so that the
constraints from $H\to VV,~\gamma\gamma,~hh$ searches can be
relaxed.

For $m_A=$ 700 GeV, the lower-left panel of Fig. \ref{bsm} shows the constraints of pre-LHC and the 125 GeV Higgs signal data require
$m_H>$ 300 GeV. The $A\to HZ$ channel can not give the further constraints on the parameter space.
The other features of the parameter space are similar to those of $m_A=600$ GeV.

\begin{figure}[tb]
 \epsfig{file=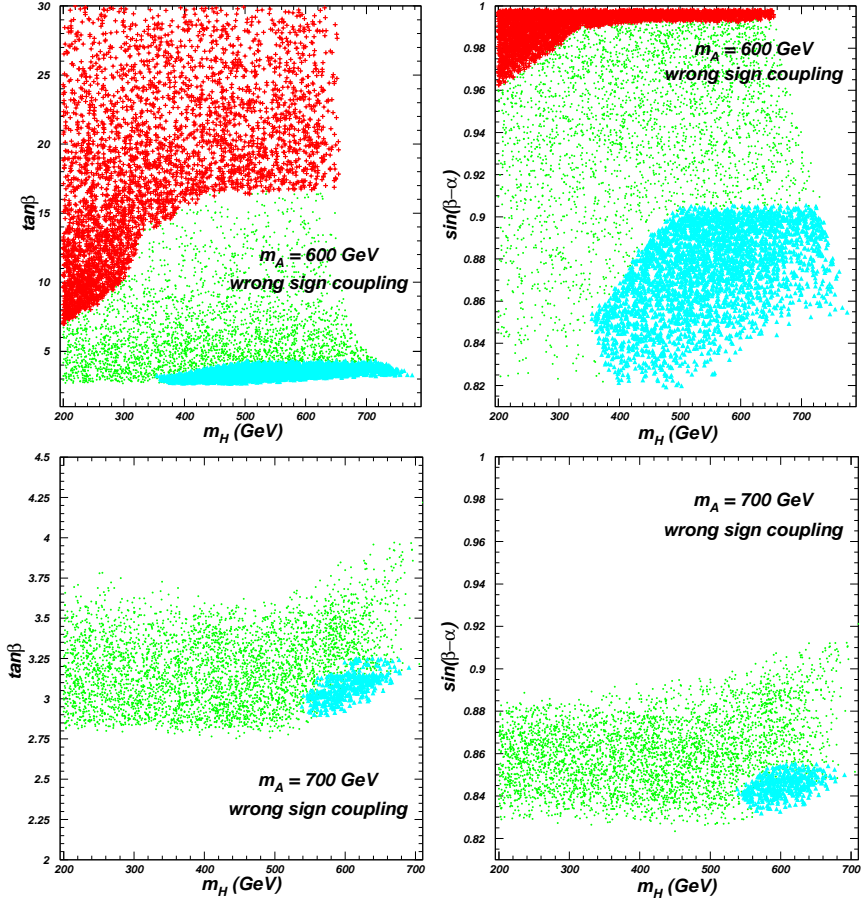,height=12cm}
\vspace{-0.5cm} \caption{The surviving samples of the
wrong sign Yukawa coupling region projected on the planes of $m_H$ versus
$\tan\beta$ and $m_H$ versus $\sin(\beta-\alpha)$. All the samples
are allowed by the constraints of pre-LHC and the
125 GeV Higgs signal data. The pluses (red) and
triangles (sky blue) are
respectively excluded by the $H/A\to \tau\bar{\tau}$ and $A\to hZ$ searches at the LHC
run-I and run-II.} \label{bw}
\end{figure}
Now we examine the parameter space in the wrong sign Yukawa coupling region for $m_A=$ 600 GeV and $m_A=$ 700 GeV.
In Fig. \ref{bw}, we project the
surviving samples on the planes of $m_H$ versus $\tan\beta$ and $m_H$ versus $\sin(\beta-\alpha)$
after imposing the constraints of pre-LHC, the 125 GeV Higgs signal data and
the searches for additional Higgses at the LHC run-I and run-II. Since the constraints of pre-LHC and the 125 GeV Higgs
signal data require $\tan\beta$ to be much larger than 2 in the wrong sign Yukawa coupling region, the $H\to VV,~\gamma\gamma,~hh$
and $A\to HZ$ channels can not give the further constraints on the parameter space.

For $m_A=$ 600 GeV, the $b\bar{b}\to H \to \tau\bar{\tau}$ channel can impose the upper bounds
on $\tan\beta$ and $\sin(\beta-\alpha)$. For example, $\tan\beta<$ 7.0 and $\sin(\beta-\alpha)<$ 0.96
for $m_H=$ 200 GeV, $\tan\beta<$ 10.0 and $\sin(\beta-\alpha)<$ 0.98
for $m_H=$ 300 GeV, $\tan\beta<$ 16.0 and $\sin(\beta-\alpha)<$ 0.99
for $m_H=$ 600 GeV. All the samples in the range of $m_H<$ 350 GeV can survive from the constraints
of $A\to hZ$ channel. In such range of $m_H$, the $A\to HZ$ mode can enhance the total width of $A$, and lead
the branching ratio of $A\to hZ$ to be suppressed. For $m_H>$ 350 GeV, the $A\to hZ$ channel can
exclude most of samples in the range of $\tan\beta<4.5$ and $\sin(\beta-\alpha)<0.905$.

For $m_A=700$ GeV, the constraints of pre-LHC and the 125 GeV Higgs
signal data require $\tan\beta<4$ and $\sin(\beta-\alpha)<$ 0.92. Therefore,
the $b\bar{b}\to H \to \tau\bar{\tau}$ channel can not give the further constraints on the parameter space.
The $A \to hZ$ channel can exclude most of samples in the range of $m_H>540$ GeV, $\tan\beta<3.25$ and
 $\sin(\beta-\alpha)<0.855$.

\section{Conclusion}
We examine the parameter space of 2HDM of type-II after imposing the
relevant theoretical and experimental constraints from the precision
electroweak data, $B$-meson decays, $R_b$, the 125 GeV Higgs signal
data, and the $H/A\to \tau\bar{\tau},~\gamma\gamma$, $H\to
WW,~ZZ,~hh,~AZ$, $A\to hZ,~HZ$ searches at the LHC run I and run II.
We obtain the following observables:

(i) Status of CP-odd Higgs $A$. Due to the constraints of theory and oblique parameters, for $m_H$ is around 600 GeV $\sim$
700 GeV, the $A$ is allowed to have a wide range of mass, including the low mass.
In the SM-like Higgs coupling region of the 125 GeV Higgs,
the $A\to hZ,~\gamma\gamma,~\tau\bar{\tau}$ channels can exclude 140 GeV $<m_A<$ 350 GeV.
For $m_H=600$ GeV, the $H\to AZ$ can exclude most of samples
in the range of $m_A<200$ GeV. The $b\bar{b}\to A/H\to \tau\bar{\tau}$ can
impose the upper limits on $\tan\beta$ and $\mid\sin(\beta-\alpha)\mid$ in the large mass range.
The parameter space of $\tan\beta<$ 2, 0.99$\leq\sin(\beta-\alpha)\leq$ 1 and -1$\leq\sin(\beta-\alpha)\leq$-0.998 are allowed
for $m_H$=600 GeV and 350 GeV $<m_A<$ 540 GeV as well as $m_H$=700 GeV and 350 GeV $<m_A<$ 640 GeV.

In the wrong sign Yukawa coupling region of the 125 GeV Higgs,
for $m_H=$ 600 GeV and 280 GeV $<m_A<700$ GeV, the $\tan\beta$ and $\sin(\beta-\alpha)$ can be imposed the
upper bounds by the $b\bar{b}\to A \to \tau\bar{\tau}$ channel
and the lower bounds by the $A\to hZ$ channel. The $b\bar{b}\to A \to \tau\bar{\tau}$ channel
can exclude most of samples in the range of
 $m_A<$ 200 GeV except for a very narrow band of $m_A$ around 100 GeV.
Compared to the case of $m_H=600$ GeV, for $m_H=700$ GeV,
320 GeV $< m_A<$ 500 GeV is excluded, and the allowed parameter space is sizably narrowed since
the constraints of pre-LHC and the 125 GeV Higgs signal data require $\tan\beta<5.5$ and $\sin(\beta-\alpha)<$ 0.95.

(ii) Status of heavy CP-even Higgs $H$.  For $m_A$ is around 600 GeV, the $H$ is allowed to have a wide rang of mass.
 In the SM-like Higgs coupling region of the 125 GeV Higgs,
the $b\bar{b}\to H/A\to \tau\bar{\tau}$ searches give the upper bound of $\tan\beta$, such
as $\tan\beta<6$, 10 and 15 for $m_H=200$ GeV, 320 GeV and 600 GeV.
The $H\to \tau\bar{\tau},~WW,~ZZ,~\gamma\gamma,~hh$ channels require
$\tan\beta>$ 2.5 for $m_H<$ 380 GeV. For $m_A=$ 600 GeV, the $A\to
HZ$ channel can exclude most of samples in the range of $m_H<$ 270
GeV. For the proper $\tan\beta$ and $\sin(\beta-\alpha)$, $m_H$ is
allowed to be as low as 200 GeV for $m_A=600$ GeV and 300 GeV for
$m_A=700$ GeV.

In the wrong sign Yukawa coupling region of the 125 GeV Higgs,
the $b\bar{b}\to H \to \tau\bar{\tau}$ channel can impose the upper bounds
on $\tan\beta$ and $\sin(\beta-\alpha)$. For $m_A=$ 600 GeV, the $A\to hZ$ channel can
exclude most of samples in the range of $\tan\beta<4.5$, $\sin(\beta-\alpha)<0.905$ and $m_H>$ 350 GeV.
Compared to the case of $m_A=600$ GeV, for $m_A=700$ GeV, the allowed parameter space are sizably narrowed since
the constraints of pre-LHC and the 125 GeV Higgs signal data require $\tan\beta<4$ and $\sin(\beta-\alpha)<$ 0.92.
For the proper $\tan\beta$ and $\sin(\beta-\alpha)$, $m_H$ is allowed to be as low as 200 GeV for
both $m_A=600$ GeV and $m_A=700$ GeV.

\section*{Acknowledgment}
We thank Qing-Hong Cao and Ye Chen for helpful discussions.
 This work is supported by the National Natural Science Foundation
of China under grant No. 11575152.

\end{document}